# Supertetragonal BaSnO$_3$ induced giant ferroelectricity in SrTiO$_3$/BaSnO$_3$ superlattices

Jing Li[1], Qing Zhang[1], Karin M. Rabe[2, *] and Xiaohui Liu[1, †]

[1] *School of Physics, Shandong University, Ji'nan 250100, China*
[2] *Department of Physics and Astronomy, Rutgers University, Piscataway, New Jersey 08854, USA*

Perovskite BaSnO$_3$ has an Sn s-orbital conduction band minimum, which makes it of interest as a transparent-conducting oxide parent compound but also contraindicates the ferroelectric instability characteristic of the related compound BaTiO$_3$. In this work, we studied the effect of (001) compressive strain on BaSnO$_3$ using first-principles methods. We found that, with low compressive strain, symmetry breaking takes cubic BaSnO$_3$ to a nonpolar tetragonal state, with a first-order phase transition to a hidden highly-polarized ferroelectric supertetragonal state at about -5%. Based on the facts that the mismatch of lattice constant in experiment between BaSnO$_3$ and SrTiO$_3$ is about -5.2% and coherent growth of BaSnO$_3$ on SrTiO$_3$ has been experimentally realized for BaSnO$_3$ layers thinner than 3 unit-cells, we studied a series of SrTiO$_3$/BaSnO$_3$ superlattices with one or two unit-cells of BaSnO$_3$ and several unit-cells of SrTiO$_3$. We found that the superlattices are ferroelectric with large polarizations. We propose that the origin of ferroelectricity in the superlattices is the mechanical and electrical coupling of the BaSnO$_3$ and SrTiO$_3$ layers, with polarized supertetragonal state of BaSnO$_3$ induced by compressive-strain from the SrTiO$_3$ layers and polarization of the SrTiO$_3$ layers by the polar BaSnO$_3$ layers. Due to the distinctive electronic states in the BaSnO$_3$ layers, the realization of ferroelectricity holds promise for the design of novel electronic devices.

**Introduction**

In nature, perovskite ferroelectric oxides are relatively rare due to rotation of the oxygen octahedra, which tends to suppress polar distortions [1]. The investigation of new mechanisms for ferroelectricity in bulk materials and heterostructures thus has practical importance for novel electronic devices. In particular, advances in the epitaxial growth of coherent films and heterostructures make it possible to combine two or more distinct materials to artificially tune the ferroelectric properties of the system and even couple ferroelectricity to other properties [2, 3]. For example, it was shown that in SrTiO$_3$/BaTiO$_3$ superlattices, the polarization of the BaTiO$_3$ layers is enhanced due to the compressive strain applied by the SrTiO$_3$ layers [4]. In short-period superlattices combining a ferroelectric and a metal, it has been shown that the polarization of the ferroelectric layer along the normal direction can be maintained with the ultrathin metallic layer acting as a dielectric in the normal direction [5]. Finally, combining ferroelectric and magnetic materials in superlattices has been shown to be an effective way to realize multiferroics [6, 7, 8].

Ferroelectricity in heterostructures can also be obtained by combining two or more compounds that are not individually ferroelectric [9, 10, 11, 12]. In such cases, the coupling of the compounds leads to reconstruction of electronic structures or atomic structures, resulting in ferroelectricity. One example is found in charge-order-driven ferroelectrics, such as LaVO$_3$/SrVO$_3$ superlattices [9], in which symmetry breaking due to charge ordering combines with the symmetry of the cation ordering in the superlattices, can generate ferroelectricity. However, the electron-lattice coupling tends to suppress polarization switching in this kind of ferroelectric system [13]. Another mechanism for ferroelectricity in heterostructures of nonferroelectric components is stabilization of a competing ferroelectric phase of one of the nonferroelectric components. For example, in perovskite oxides, the suppression of octahedral rotations could favor a polar distortion. Indeed, it has been theoretically predicted and experimentally confirmed that nonpolar CaTiO$_3$ can be stabilized to a polarized BiFeO$_3$-type phase in superlattices [10, 14] and thin films [15, 16, 17].

BaSnO$_3$ is a cubic perovskite that has recently been of great interest as a parent compound that can be doped to produce transparent conducting oxides with high mobility [18]. This is due to its isolated and dispersive s orbital conduction band minimum [19] giving mobility about two orders of magnitude higher than perovskite oxides with d orbital conduction band minima [20, 21]. However, this also contraindicates the ferroelectric instability characteristic of the related compound BaTiO$_3$, in which the d-orbital character of the conduction band minimum plays a key role.

As described in more details below, our first-principles calculations show that epitaxially strained BaSnO$_3$ undergoes a structural transition from its tetragonally strained nonpolar ground state structure to a ferroelectric highly-polar supertetragonal state at a critical strain of about -5%. Previous studies have reported such supertetragonal structures in PbTiO$_3$ [22], BaTiO$_3$ [23] and BiFeO$_3$ [24, 25]. The mismatch of experimental lattice constant values between SrTiO$_3$ and BaSnO$_3$ is -5.2%. While this is quite large, recent experimental study has shown that coherent growth of BaSnO$_3$ on SrTiO$_3$ substrate could be realized within a critical thickness of about three unit cells [26], demonstrating the feasibility of the growth of SrTiO$_3$ / BaSnO$_3$ superlattices with ultrathin BaSnO$_3$ layers.

In this paper, we use first-principles methods to investigate the possibility of stabilization of the supertetragonal phase of BaSnO$_3$ and resulting giant ferroelectricity in (SrTiO$_3$)$_m$/(BaSnO$_3$)$_n$ superlattices using first-principles methods. We find that with $n = 1$ or 2, the supertetragonal phase is stabilized by the strain due to the lattice mismatch with SrTiO$_3$, and that even for $m$ as large as 12, the superlattices have spontaneous polarization larger than that of tetragonal BaTiO$_3$, thanks to the high polarizability of the SrTiO$_3$ layers.

First-principles density functional theory (DFT) calculations are performed using the plane-wave pseudopotential code Quantum ESPRESSO [27]. In the calculations, the electron wave functions are expanded in a plane-wave basis set limited by a cutoff energy of 550 eV, and the exchange and correlation effects are treated within the generalized gradient approximation (GGA). Periodic boundary conditions are employed in all three directions. The calculated in-plane lattice constant of cubic $SrTiO_3$ is 3.94 Å (this overestimate compared with experimental $a$ = 3.905 Å is typical of GGA results). $BaSnO_3$ is cubic with a lattice constant of 4.19 Å in our calculation (compare the experimental value 4.116 Å [28]). The supercells are constructed by stacking the two structural unit-cells, namely $SrTiO_3$ and $BaSnO_3$, along the [001] direction (which we consider the z-axis. The $SrTiO_3$ layers are $TiO_2$ terminated and the $BaSnO_3$ layers are BaO terminated, giving symmetric supercells. To simulate coherent epitaxial growth on a (001) oriented substrate of $SrTiO_3$, we constrain the in-plane structure of each bulk material component of the heterostructure to a 1×1 cubic perovskite cell with lattice constant a =3.94 Å consistent with the calculated lattice constant of cubic $SrTiO_3$ and perform full relaxation of the internal z-coordinates and tetragonal out-of-plane lattice constant c. For the structural relaxation, the Brillouin zone is sampled with a 6×6×1 Monkhorst-Pack k-point mesh using the force convergence criterion of 10 meV/Å.

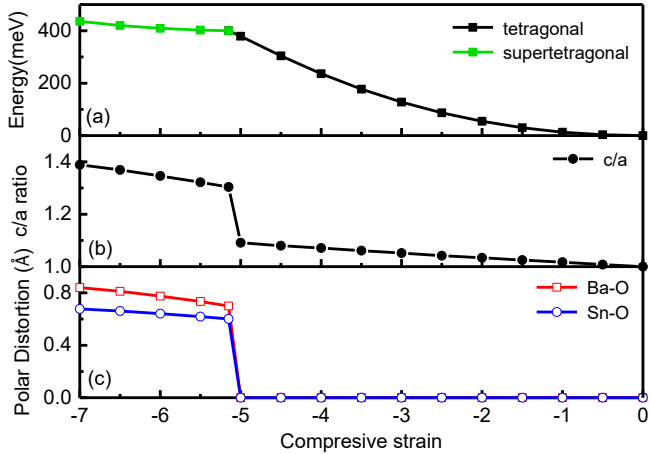

Fig. 1. (a) Total energy, (b) c/a ratio and (c) polar distortion, i.e. the relative displacements of Ba-O and Sn-O as a function of stain in bulk $BaSnO_3$. The calculated energy of cubic $BaSnO_3$ is set to be zero. Negative values of the horizontal axis mean that the strain is compressive.

First of all, to show the effect of compressive strain on stabilizing the supertetragonal state of $BaSnO_3$, we considered the effect of compressive epitaxial strain on bulk $BaSnO_3$ by constraining the in-place lattice constant and relaxing $c$. As shown in Fig. 1(b), as the compressive epitaxial strain constraint drives the cubic $BaSnO_3$ structure to tetragonal, with the c/a value increasing with the magnitude of compressive strain. As shown by Fig. 1(c), for compressive strains smaller than about -5%, the tetragonal phase is still centrosymmetric, with no polar distortion. At a critical value of -5%, there is a jump in both the c/a ratio and the polar distortion, resulting in a distinct highly-polar supertetragonal phase (Fig. 1(b) and 1(c)). To be precise, the c/a jumps from 1.08 in the nonpolar tetragonal state to a value larger than 1.25 in the super-tetragonal state. From Fig. 1 (a), we can see that, the energy profile has a cusp characteristic of a first-order transition at around -5% compressive strain.

The spontaneous polarization of the supertetraonal state is very large, comparable to the largest polarizations seen for ferroelectric perovskites. Specifically, our first-principles calculation based on the modern theory of polarization [29, 30] shows that the polarization at -5.5 % compressive strain is 86 $\mu C/cm^2$. The relative positions of atoms in the unit cell of supertetragonal $BaSnO_3$ are shown in Fig.2(a).

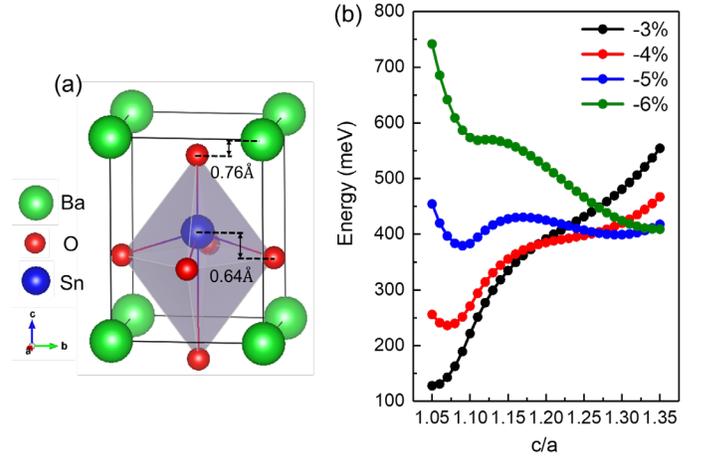

Fig. 2. (a) Atomic configuration of $BaSnO_3$ with compressive strain of -5.5%; (b) Energy profile as a function of c/a ratio for four fixed in-plane strain values. The energy of cubic $BaSnO_3$ is taken as the zero of energy.

To give a better understanding of the competition between nonpolar tetragonal and highly-polar supertetragonal states under compressive strain, in Fig. 2 (b) we show the energy profile as a function of c/a ratio for several cases of fixed compressive strain: -3%, -4%, -5% and -6%. At -3%, there is a single minimum corresponding to the nonpolar tetragonal state. At -4%, a second local minimum at higher c/a is clearly emerging. At -5%, we see two local minima correspond to the nonpolar tetragonal and highly-polar supertetragonal states of $BaSnO_3$, and at -6%, we see that the local minimum corresponding to the nonpolar tetragonal state is on the verge of disappearing. So, above about -5%, $BaSnO_3$ is stable in the supertetragonal phase.

As reported in experiments [28], for coherent growth of $BaSnO_3$ on $SrTiO_3$, the thickness of $BaSnO_3$ should be less than 1.2 nanometers, which is about 3 unit-cells. Therefore, we studied a series of $(SrTiO_3)_m/(BaSnO_3)_n$ supercells, with the thickness of the BaSnO₃ layer fixed at $n = 1$ or $n = 2$ unit-cells and the thickness of the SrTiO₃ layer $m$ to be varied. In our calculations, both interfaces of the constituent layers are TiO₂/BaO as shown in Fig. 3(a), where the atomic arrangement in the supercell with $m = 8$ and $n = 2$ is shown.

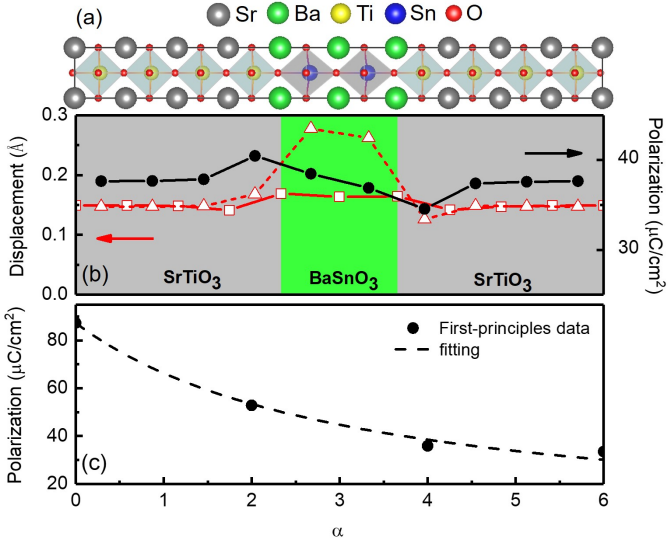

Fig. 3. (a) Supercell of a (SrTiO₃)$_m$/ (BaSnO₃)$_n$ superlattice with $m = 8$ and $n = 2$. (b) Relative displacement along the z direction between metal cation and anion (oxygen) in each atomic layer. Square symbols correspond to the AO layers (A=Sr or Ba) and triangular symbols correspond to the BO₂ layers (B= Ti or Sn). The black circles show the local polarization in each local unit cell layer computed from the Born effective charges method (Equation 1); (c) Polarization computed from first-principles (black dots) and polarization fitted using Eq. (2) as a function of $\alpha = m/n$ for four superlattices with fixed $n = 2$ and $m = 0$ (pure strained BaSnO₃), 4, 8 and 12.

It is remarkable to find that, even though neither SrTiO₃ nor BaSnO₃ is ferroelectric in bulk, large polarization is exhibited in (SrTiO₃)$_m$/(BaSnO₃)$_n$ superlattices even for large $m$. To understand this, we consider $m = 8$. The relative displacements between cations and anions of each atomic layer are shown in Fig. 3 (b). Compared with the polar displacements of the supertetragonal BaSnO₃ compound as shown in Fig. 2 (a), the magnitude of polar displacements in the superlattice are reduced, but still substantial. To estimate the local polarization for each unit-cell layer of the supercell, we use a model based on Born effective charges [31], writing the local polarization $P(z)$ in each local unit cell, defined by A atoms at the corners,

$$P(z) = \frac{e}{\Omega}\sum_{i=1}^{N} Z_i^* \delta z_i \quad (1)$$

Here $\Omega$ is the volume of the local unit cell, $N$ is the number of atoms in the unit cell, $\delta z_i$ is the displacement of the $i$th atom away from its position in the centrosymmetric structure and $Z_i^*$ is the Born effective charge of the $i$th atom, calculated in the strained bulk crystal. All polarizations are along the [001] direction. Our calculated $Z_i^*$ values are 2.55, 7.35, 2.75 and 4.46 for Sr, Ti, Ba and Sn, respectively. For oxygen, there are two different positions, in the BaO layer and SnO₂ layer, and the corresponding values of $Z_i^*$ are -3.58 and -1.64 respectively. The calculated local polarization is shown in Fig. 3 (b). We can see that the whole supercell is polarized with local polarization being fairly uniform. In the BaSnO₃ layer, the polarization is reduced compared with the pure supertetragonal state, while the SrTiO₃ layer is polarized. This is expected from electrostatic energy considerations, and is characteristic of superlattices of SrTiO₃ with ferroelectric perovskite oxides like BaTiO₃ and PbTiO₃ [4]. In SrTiO₃/BaTiO₃ superlattices, both theoretical and experimental studies have shown that the polarization originates from the intrinsic polarization of the tetragonal BaTiO₃. This polarization is increased by the compressive strain due to the lattice mismatch with SrTiO₃, and reduced to match the polarization of the SrTiO₃ layer. For some superlattices, this can result in a polarization larger than that of bulk BaTiO₃ [4]. Here, in the SrTiO₃/BaSnO₃ superlattice, the compressive strain due to the lattice mismatch with SrTiO₃ drives a first-order phase transition from the nonpolar tetragonal phase to the highly polarized supertetragonal phase of BaSnO₃, while the polarization matching with SrTiO₃ leads to a reduction of the polarization of the BaSnO₃ layer and polarization of the SrTiO₃ layer.

We now consider the trends in the polarization in more details. The relatively uniform local polarization minimizes the electrostatic energy associated with the polarization charge $\nabla \cdot \mathbf{P}$. In Fig. 3 (c), we show the polarization as a function of the thickness $m$. The calculated polar displacements for various $m$ are shown in the Supplemental Material Fig. 1S. The filled circles indicate the calculated polarization as a function of $\alpha = m/n$ for $n = 2$. The dashed curve is obtained by fitting the following expression

$$\bar{P} = \frac{P_0}{1+\alpha(\varepsilon_b/\varepsilon_s)} \quad (2)$$

obtained from macroscopic electrostatics, where $P_0$ is the polarization of bulk BaSnO₃ under the compressive strain being applied by SrTiO₃; $\alpha$ is the ratio of the thickness $n$ of the BaSnO₃ layer and the thickness $m$ of the SrTiO₃ layer. $\varepsilon_b$ and $\varepsilon_s$ are the dielectric constants of BaSnO₃ and SrTiO₃, respectively. To get the best fit, the values of $\varepsilon_b/\varepsilon_s$ are 0.38 and 0.42 for $m=2$ and $m=1$ respectively. We could see that, the fitting curve by the equation matches the calculated data remarkably well. Even for long period superlattice with $m=12$, (SrTiO₃)₁₂/(BaSnO₃)₂, the polarization is more than 30 μC/cm², which is even larger than the polarization of bulk ferroelectric BaTiO₃.

Calculations for (SrTiO₃)$_m$/(BaSnO₃)₁ superlattices show similar results, even with only one unit-cell of BaSnO₃. This is shown in the Supplemental Material Fig. 2S and 3S.

The ferroelectric SrTiO₃/BaSnO₃ superlattice has important differences from well-studied ferroelectric superlattices like SrTiO₃/BaTiO₃, SrTiO₃/PbTiO₃, PbTiO₃/BaTiO₃ where at least one compound is ferroelectric. In this case, the strain in the superlattice drives a first-order phase transition of the BaSnO₃ layer from the nonpolar tetragonal phase to the supertetragonal phase. Remarkably, this transition occurs for $BaSnO_3$ layer thickness as low as a single unit cell. Another important difference is the polarized oxide $BaSnO_3$ with s orbital conduction band which is high mobility unlike the localized d orbitals in other ferroelectric superlattices. The details of the electronic properties of this structure are the subject of ongoing studies.

We should note that the calculated strain provided by $SrTiO_3$ is only slightly larger than the calculated critical strain for the phase transition of $BaSnO_3$ from tetragonal to supertetragonal. Therefore, in experiments, it is not guaranteed that the strain will be enough to produce the polarized superlattice state. Thermal fluctuations and crystal imperfections may also affect the polarized supertetragonal state in the $SrTiO_3$/$BaSnO_3$ superlattices. So, we look for ways to increase the relative stability of the supertetragonal phase of $BaSnO_3$. One way is to grow the superlattices on substrates which provide more compressive strain than $SrTiO_3$. Another way is to reduce the critical strain for the supertetragonal transition by compositional substitution in $BaSnO_3$. A third way is to favor the polarization by applying an electric field to the superlattice. This would, in effect, make the superlattices antiferroelectric, with the high polarization state achieved at nonzero electric field in a double butterfly loop, accompanied by a large effective piezoelectric response [32].

In conclusion, we propose that a novel artificial ferroelectric can be obtained by forming a superlattice of two paraelectric compounds, $BaSnO_3$ and $SrTiO_3$ [4, 33, 34]. The origin of the ferroelectricity is the stabilization of polarized supertetragonal phase of $BaSnO_3$ by the compressive epitaxial strain applied by the $SrTiO_3$ layer, with the electrostatic effect resulting in polarization of the whole system. We performed calculations on a series of $(SrTiO_3)_m/(BaSnO_3)_n$ superlattices to show that, even for a single layer of $BaSnO_3$, the value of the spontaneous polarization is comparable to that of conventional ferroelectrics. The recently reported coherent growth of an ultrathin $BaSnO_3$ layer on $SrTiO_3$ suggests that these systems could be experiementially realized. We hope that this result for stabilization of the ferroelectric phase of $BaSnO_3$ could stimulate further studies on ferroelectricity of perovskite oxides with s orbitals. Due to the distinctive electronic states in the $BaSnO_3$ layers, the realization of ferroelectricity holds promise for the design of novel electronic devices.

The work of X. L. was supported by the National Natural Science Foundation of the People's Republic of China (Grants 11974211) and Qilu Young Scholar Program of Shandong University. The work of K. M. R. was supported by the Office of Naval Research through grant N00014-21-1-2107. We acknowledge useful discussions with Charles Ahn, Frederick Walker, and Yeongjae Shin.

# Supertetragonal BaSnO$_3$ induced giant ferroelectricity in SrTiO$_3$/BaSnO$_3$ superlattices

Jing Li[1], Qing Zhang[1], Karin M. Rabe[2], Xiaohui Liu[1]

[1] School of Physics, Shandong University, Ji'nan 250100, China

[2] department of Physics and Astronomy, Rutgers University, Piscataway, New Jersey 08854, USA

## A. Superlattice (SrTiO$_3$)$_m$/(BaSnO$_3$)$_n$ with n=2

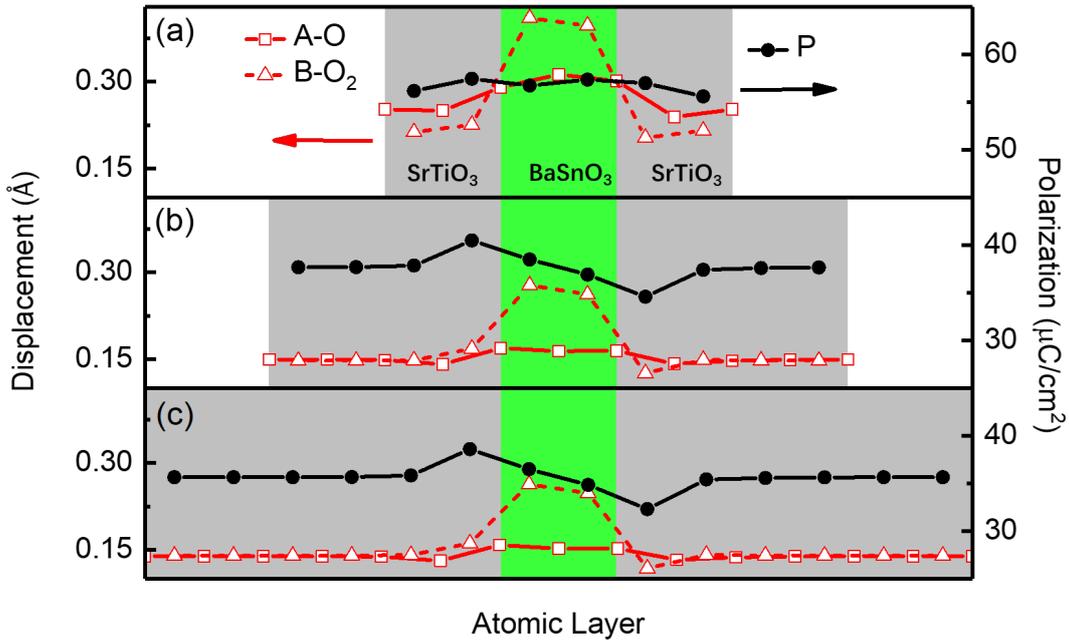

Fig. 1S. Relative displacement along the z direction between metal cation and anion (oxygen) in each atomic layer for superlattice (SrTiO$_3$)$_m$/(BaSnO$_3$)$_2$ with fixed $n = 2$ and (a) $m = 4$, (b) $m = 8$, (c) $m = 12$. Square symbols correspond to the AO layers (A=Sr or Ba) and triangular symbols correspond to the BO$_2$ layers (B= Ti or Sn). The black circles show the local polarization in each local unit cell layer computed from the Born effective charges method.

We calculated the polar displacements of each atomic layer of the (SrTiO$_3$)$_m$/(BaSnO$_3$)$_2$ superlattice with the thickness of BaSnO$_3$ layer fixed to be $n = 2$, and the thickness of SrTiO$_3$ layer varies with $m = 4$, $m = 8$ and $m = 12$. The results are shown by the red curves in Fig. 1S (a), (b) and (c) respectively. Fig. 1S (b) is the same as Fig. 3 (a).

## B. Superlattice $(SrTiO_3)_m/(BaSnO_3)_n$ with $n = 1$

We calculated the polar displacements of each atomic layer of the $(SrTiO_3)_m/(BaSnO_3)_n$ superlattice with the thickness of $BaSnO_3$ layer fixed to be $n = 1$, and the thickness of $SrTiO_3$ layer varies with $m = 2$, $m = 4$ and $m = 6$. The results are shown by the red curves in Fig. 2S (a), (b) and (c) respectively.

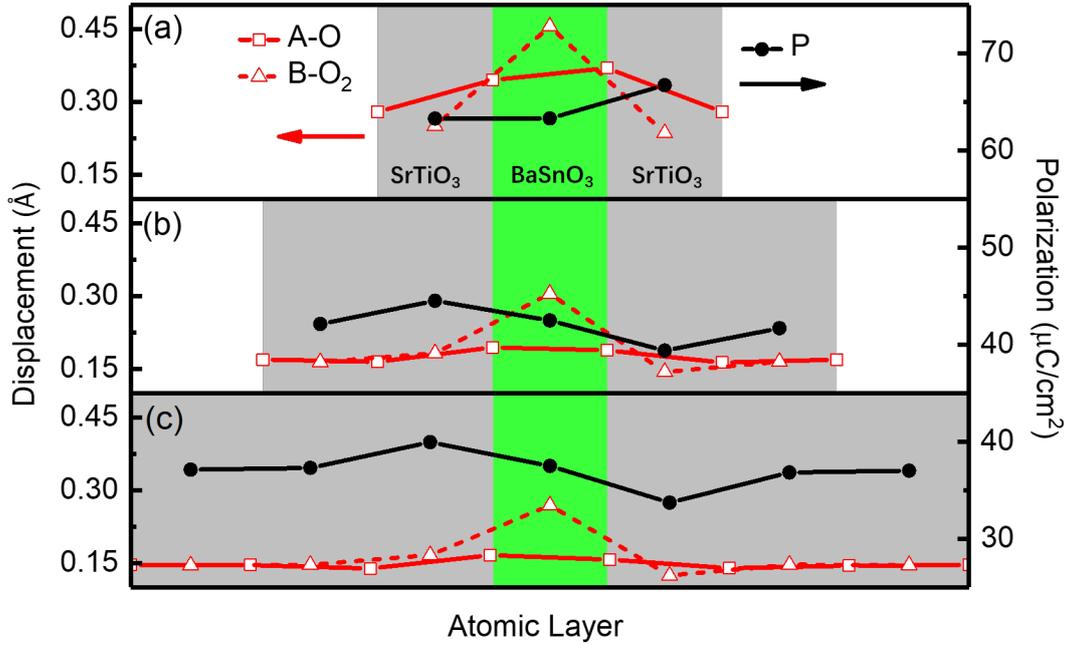

Fig. 2S. Relative displacement along the z direction between metal cation and anion (oxygen) in each atomic layer for superlattice $(SrTiO_3)_m/(BaSnO_3)_1$ with fixed $n = 1$ and (a) $m = 2$, (b) $m = 4$, (c) $m = 6$. Square symbols correspond to the AO layers (A=Sr or Ba) and triangular symbols correspond to the $BO_2$ layers (B= Ti or Sn). The black circles show the local polarization in each local unit cell layer computed from the Born effective charges method.

Fig. 3S (b) shows the polar displacement of each layer of a superlattice $(SrTiO_3)_4/(BaSnO_3)_1$ with $n = 1$ an $m = 4$ which is the same as Fig. 2S (b). The atomic structure of it is shown by the cartoon Fig. 3S (a). To show the effects of the thickness of $SrTiO_3$ layer on the $(SrTiO_3)_m/(BaSnO_3)_1$ superlattice, we polt the tendency of the polarization as a function of thickness of $m$ with fixed $n = 1$ as shown by Fig. 3S (c). We could see that the calculated polarization matchs well with the fitting result by Born effective charge method.

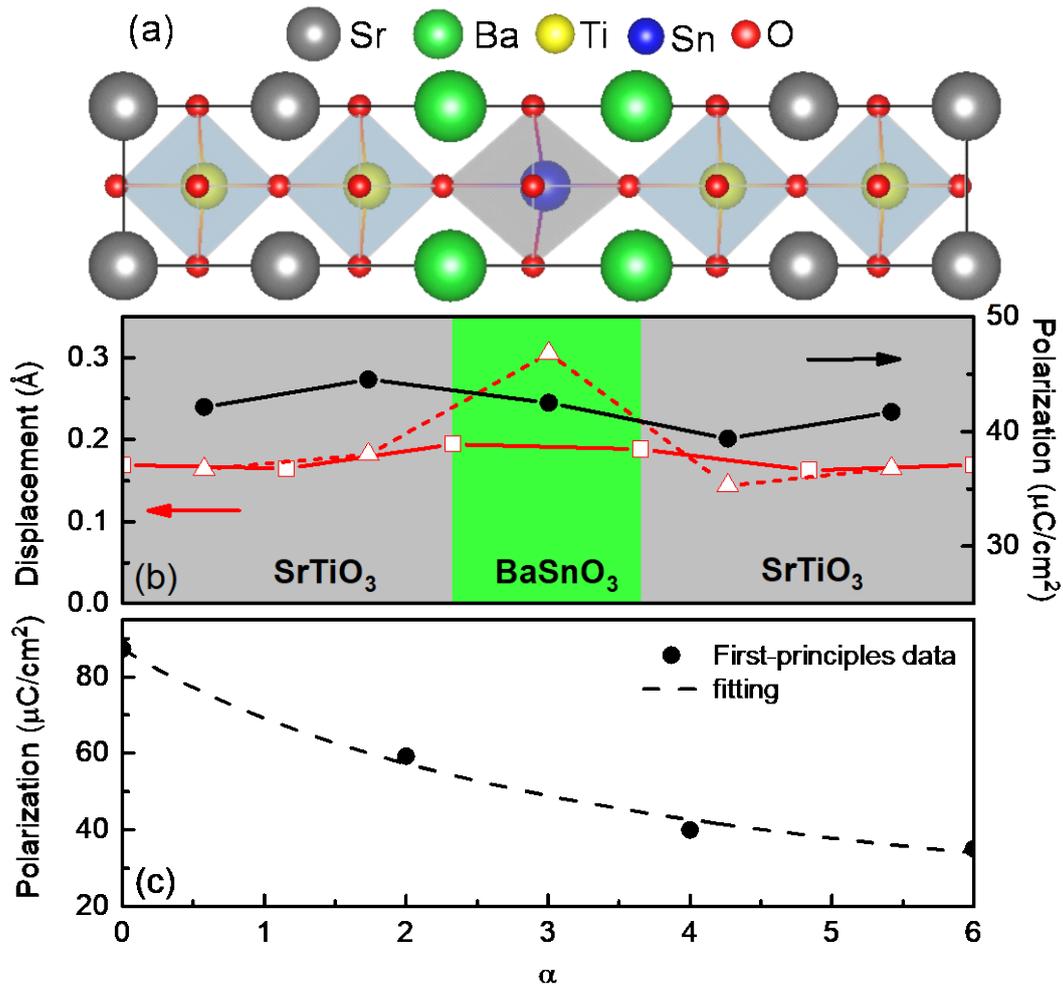

Fig. 3S. (a) Supercell of a (SrTiO$_3$)$_m$/ (BaSnO$_3$)$_n$ superlattice with $m = 4$ and $n = 1$. (b) Relative displacement along the z direction between metal cation and anion (oxygen) in each atomic layer. Square symbols correspond to the AO layers (A=Sr or Ba) and triangular symbols correspond to the BO$_2$ layers (B= Ti or Sn). The black circles show the local polarization in each local unit cell layer computed from the Born effective charges method; (c) Polarization computed from first principless (black dots) and polarization fitted using Eq. (2) as a function of $\alpha = m/n$ for four superlattices with fixed $n = 2$ and $m = 0$ (pure strained BaSnO$_3$), 2, 4 and 6.